\documentstyle[amstex,epsfig, psfig, psfrag, referee, graphics]{mn}

\def\part#1#2{{{\partial #1} \over {\partial #2}}}

\def\rr{{\bf r}}

\def\II{{\bf I}}
\def\CC{{\bf C}}
\def\DD{{\bf D}}
\def\RR{{\bf R}}
\def\rr{{\bf r}}
\def\FF{{\bf F}}
\def\SS{{\bf S}}
\def\ee{{\bf e}}
\def\nn{{\bf n}}

\def\ll{{\bf l}}
\def\mm{{\bf m}}

\begin{document}
\title{The optical polarization of spiral galaxies}

\author[ J.F.L. Simmons and E. Audit]
{ John F.L. Simmons$^{1}$, Edouard Audit$^{2,3}$ \\
  $^{1}$ Department of Physics and Astronomy, University of Glasgow
  Glasgow G12 8QQ - U.K\\
  $^{2}$ Service D'Astrophysique, CEA/Saclay, Ormes des Merisiers - 91191 Gif-sur-Yvette - France\\
  $^{3}$ Laboratoire d'Astrophysique Extragalactique et de Cosmologie
   -  CNRS URA 173 - \\
  Observatoire de Meudon - 5, Place Jules Jansen - 92
  195 Meudon - France}

\date{Accepted 199-
      Received 199-
      in original form 199-}

\maketitle
\begin{abstract}
Scattering of starlight  by dust, molecules and  electrons in spiral galaxies
will produce a modification of the direct intensity and a polarization in the
observed light. We treat the case where the distribution of scatterers can be
considered to be optically thin, and derive semi-analytic expressions for the
resolved  intensity and polarized intensity  for  Thomson, Rayleigh, and more
general scattering mechanisms. These  expressions are applied to a parametric
model spiral galaxies.  It is further  shown that in  the case of Thomson and
Rayleigh scattering, and   when  scatterers and  stars are  distributed  with
rotational  symmetry, the total   polarized flux depends  on the inclination,
$i$, of the galactic axis to the line of sight according to a simple $\sin ^2
i$ law.   This generalises the well  known result for pointlike and spherical
light sources. By using a method based  on spherical harmonics, we generalise
this law  for more general mechanisms, and  show that  to good approximation,
the $\sin ^2 i$ law still holds for the class of models considered.
\end{abstract}
\section{Introduction}

The central discs of spiral galaxies are  known to contain substantial
quantities of dust, molecules  and free electrons.  The scattering  of
light  from stars, which  are  largely confined to   the disc and  the
bulge,  by these dust  particles,  molecules and electrons can broadly
explain the  observed optical   intensity and polarization   of spiral
galaxies.

Maps of optical  intensity of a large number  of  spiral galaxies have
been obtained  observationally, and  polarization  maps for a  smaller
number. Several galaxies have  been more or less successfully modelled
for their intensity and polarization by Monte Carlo simulations (\cite
{Wood97a}), a technique  suited  to the  optically thick  regime. Such
studies have yielded more information about the dust content of spiral
galaxies  (\cite {Byun94}),   and in some    cases have indicated  the
presence  and strength of   magnetic  fields (\cite{Draper95} ,  \cite
{Scarrott96}). Whether the galaxies are  optically thin, or  optically
thick has a crucial bearing on our understanding of galactic evolution
(\cite  {Calzetti99}) and star  formation. Similarly, the detection of
magnetic    fields   would  have  considerable  importance    for  our
understanding of   galactic   evolution. Clearly absorption  by   dust
particles and gas  will produce an attenuation of  light, and hence  a
decrease in the apparent brightness of galaxies.  For this reason such
calculations  have a  practical   bearing on  distance  estimation  of
galaxies.

Monte Carlo simulations  undoubtedly provide a  powerful technique for
understanding  the physical processes taking  place in the galaxy, but
often, because of their dependence  of a specific choice of geometries
and parameter values, can  obscure certain fundamental properties  and
relationships   that might be   obtained   through a simple   analytic
approach.

In this paper    we assume that  the   distribution of scatterers   is
optically  thin.  This  assumption allows  one to  obtain  a number of
interesting analytic results    for cases where  the   distribution of
scatterers   and of  stars  is   symmetric,  and  yields fairly simple
expressions  for the unpolarized  and polarized intensity and flux for
the more  general case in terms  of simple integrals.  In  the case of
many spiral galaxies the  assumption of  optical thinness is  probably
not unreasonable (\cite  {X99}) , (\cite  {Byun94}) but even in  those
cases where it is not expected to hold, the optically thin results can
often give a  qualitative  picture of  what is happening.   It is also
possible to  give a semi-analytic treatment of  the the case where the
galaxy  is optically    thick in absorption,  but   optically  thin in
scattering. This we shall deal with in a  future paper.  If the spiral
galaxy is considered to have  a rotational axis  of symmetry we expect
the  direction of polarization to be  along (or possible perpendicular
to) direction  of axis  of  symmetry projected perpendicularly to  the
line of  sight.  (Of course  this axis of  symmetry would be broken by
the  presence of  spiral  arms,  but  even so   would be approximately
valid.)  It  has    been   emphasised (\cite   {Audit99})   that  this
orientation of the   total integrated light polarization  of  galaxies
could play and  important role in  studies of the distribution of dark
matter  from weak  lensing,  which  has the  effect   of changing  the
orientation of the semi-major axis of  the elliptical isophotes of the
galaxy,  but leaves   the  direction of  polarization  unchanged.  The
difference between the direction of the image semi-major axis, and the
polarization direction  thus gives and  indication of the  strength of
the  lensing.     This  potentially  would   considerably  reduce  the
uncertainty  in inferring  the  mass distribution   with in  the  lens
compared  with  the  usual  weak  lensing   studies, which   take  the
orientation of the  source galaxy as unknown.   We show in  this paper
that in this symmetric and optically thin case for the case of Thomson
and Rayleigh   scattering obeys a  $\sin^2  i$ law,  where $i$  is the
inclination  of the   axis of  symmetry to  the  line  of sight.  This
generalizes   the well known result for   Thomson scattering for point
light  sources sources    \cite{Brown77}, \cite{Simmons82},  and   for
spherical extended sources \cite{Cassinelli86}  derived in the stellar
context. We further show that even for other more realistic scattering
phase    functions this  law   approximately holds  for typical galaxy
models, generalising the results of
\cite{Simmons83}.

The structure of  paper is as follows.   In section  2 we introduce  a
widely accepted parametric model for spiral galaxies.  In section 3 we
give the  basic definitions of Stokes intensities  and  fluxes and set
down the equations of radiative  transfer for the Stokes  intensities,
and specialise these to the case where the source of light is provided
by  a  distribution  of stars.   In section   \ref{optically thin}  we
discuss the optically thin case, and derive expressions for the Stokes
intensities  and fluxes   for   Rayleigh and more  general  scattering
mechanisms, and compare the  polarization maps with those obtained  by
Monte  Carlo techniques.  We  go  on to outline   a  general and  very
powerful  method for treating  the  optically  thin  case  for general
scattering mechanisms that makes use of  spherical harmonics and their
properties under     the  rotation group.   Finally   we  present  our
conclusions in section \ref{conclusions}.

\section{Model galaxies}\label{mod_gal}

In  this section we outline   the model  which   we have  adopted for  spiral
galaxies. The galaxies  are taken to  be composed of a  disc and bulge, which
contain stars and scattering dust and particles.  We take the distribution of
stars \cite{Jaffe} in the bulge to be given by

\begin{equation}
\rho_b(r) = \rho_b^o \frac{1}{(r/r_b)^2 \, (1 + r/r_b)^2}
\end{equation}
where $r$   is   the  distance to  the   galactic   centre and  $r_b$   is  a
characteristic   radius.     Following   other    studies  (\cite{Bianchi96},
\cite{Wood97b})  we use a value  of $ r_b = 1.0  \; kpc$.   We have also, for
numerical reasons, applied a cut-off of  the bulge distribution at a distance
of $2.5 \: kpc$.

The distribution of the stars in the disk is given by:
\begin{equation}
\label{disk}
\rho_d(R,Z) = \rho_d^o exp(-R/R_d) \, exp(-Z/Z_d) ,
\end{equation}
where $R$ is the distance to the  axis of symmetry  of the galaxy and $Z$ the
distance  to the  galactic plane. The  typical  scale length for  the disc is
taken  to be $R_d = 4  \: kpc$ with   a cut-off at   $20 \: kpc$. The typical
thickness is $Z_d = 0.5 \: kpc$ with a cut-off at $2.5 \: kpc$.

The  distribution of the   scattering particles (dust  or  electrons) is also
taken to be of the same form as equation (\ref{disk}), i.e.
\begin{equation}\label{dust}
n(R,Z) = n_0 exp(-R/R_g) \, exp(-Z/Z_g)
\end{equation}

The dust disc is assumed  to have the same  radial extend as the stellar disc
(i.e. $R_g = R_d = 4 \: kpc$, with a $20 \: kpc$ cut-off), but to be thinner:
$Z_g =  0.25 \: kpc$ with a  cut-off at $1.25 \: kpc$.   The total content of
dust is normalized by stipulating that the total  optical depth of the galaxy
along its axis of symmetry. With the form  given by equation \ref{dust}, this
is given by $\tau  = Z_g n_0 \sigma$,  where $\sigma$ is the total scattering
cross section. In our numerical models we shall  take $\tau = 0.05$. Since we
are in the optically thin approximation, results for other optical depths can
be obtained with a linear scaling.

\section{Equation of radiative transfer for the Stokes parameters}
\label{transfer}

In this  section we set down the  equations of radiative  transfer, and apply
them to  our  model  spiral  galaxy.  We  are essentially interested   in the
(asymptotic)  Stokes intensities and the polarized  and unpolarized fluxes as
measured by a terrestrial observer.

The brightness, and the degree of  linear and circular polarization of
a radiation field is  be described by  the Stokes intensities, denoted
by $I, Q, U, V$. $I$ is  the unpolarized intensity, $Q$ the difference
in   intensity  measured by a   polaroid  aligned in two perpendicular
directions, say $x$ and $y$,  $U$ the difference in intensity measured
in two perpendicular direction at $45 ^0$ to the $x$ and $y$ axis, and
$V$ the circular polarization.  $I,  Q, U, V$   are each functions  of
position, $\rr$, and direction $\nn$. For convenience we sometimes use
the notation   $I_1 = I$, $I_2= Q$,    $I_3= U$, and   $I_4 =  V$, and
introduce the `vector', ${\bf  I} = (I_1  ,I_2  ,I_3 ,I_4)^T$ for  the
Stokes intensities. The vectorial  Stokes  fluxes are defined as   the
integrated intensities, viz.

\begin{equation}\label{flux definition}
{\bf F}_i  = \int_{4\pi}  I_i \ {\bf n }' d\Omega_{\bf n '}.
\end{equation}
The flux,  corresponding to polarization  state $i$, across  a surface
oriented in  direction ${\bf n  }$ will  then  be ${\bf n }\cdot  {\bf
F}_i$. We shall be   interested in the  flux across  a surface  in the
direction of  the observer.  This   is  a scalar quantity.  We   shall
sometimes use   the obvious notation  $F_1  = F_I$,  $F_2= F_Q$, $F_3=
F_U$, and $F_4 = F_V$, and ${ \bf  F} = (F_1  ,F_2 ,F_3 ,F_4)^T$ .  To
simplify   notation, we shall    implicitly  assume  that the   Stokes
intensities can  depend on   wavelength   rather than  use  a   lambda
subscript. The degree of linear polarization  is given by $\sqrt{F^2_Q
+ F^2_U }/F_I$, and the position angle  of the polarization by $1/2 \, 
{\mbox{arctan}} F_Q / F_U $.

The direct source of light is supplied by the stars, and this light is
scattered by electrons, or absorbed, scattered or emitted by molecules
and dust.  Emission from  galactic clouds could easily be incorporated
into the  analysis, as can  absorption, but  for simplicity we  ignore
these. We assume that the  starlight is unpolarized, although this too
could be included in the present formalism.   Throughout this paper we
take the galaxy to be optically thin.   This assumption appears to the
valid for most spiral  galaxies \cite{X99},\cite{Bosma92}, although in
some cases might break down \cite{Scarrott96}.

Optically  thick cases have previously  been treated using Monte Carlo
techniques \cite{Bianchi96}, \cite{Wood97b}, but  the purpose of  this
paper is  to  treat the simplified, though  realistic,  problem by the
simplest  techniques. The case where  the scatterers  are electrons or
Rayleigh  scatterers is     largely   amenable  to   simple   analytic
treatment.  In all  our calculations we  assume, fairly realistically,
that the stars  are unpolarized  sources. Usually we  take a  model in
which both stars and dust are continuously distributed with rotational
symmetry about an axis of symmetry of the galaxy.

The equation of radiative transfer in its full generality may be
written
\begin{equation}\label{radiative transfer}
 \frac{1}{c} \part{\II (\rr , \nn , t ) }{t} + \nn .\nabla \, \II (\rr , \nn , t) =
 \CC (\rr , \nn , t ) - \DD (\rr ,\nn , t).
\end{equation}
Equation (\ref{radiative transfer}) is of course a set of four
partial differential equations for the four Stokes intensities $I,
Q, U, V$. We consider only time independent solutions, and thus all
quantities will be taken to be independent of $t$. $\CC (\rr ,
\nn)$ is the energy of the corresponding polarization state
scattered or emitted into direction $\nn $ per steradian per unit
time per unit volume, and $\DD (\rr , \nn )$ is the corresponding
energy removed per steradian per unit time per unit volume at
position $\rr$. Generally $\DD (\rr , \nn )$ will be linear in $\II
(\rr , \nn )$. It is convenient to write

\begin{equation}\label{creation}
\CC (\rr , \nn ) = \CC_{\rm stars} (\rr , \nn )+ \CC _{\rm emiss}(\rr , \nn ) +
\CC_{\rm scatt} (\rr , \nn )
\end{equation}
where $\CC _{\rm stars}(\rr  , \nn )$ is  the contribution from stellar light
sources, which  we  would   expect to  be   isotropic  (i.e.  independent  of
$\nn$). $\CC _{\rm emiss}  (\rr , \nn   )$ is the contribution from  emission
processes.  Stimulated emission  would depend  linearly  on $\II (\rr ,\nn )$
(such a  term is  easily incorporated  into the  following analysis by simple
incorporating it in the term $\DD (\rr ,  \nn )$). Thermal emission should be
isotropic and thus can be incorporated into the stellar term. Here however we
shall for simplicity  ignore  $\CC  _{\rm  emiss}(\rr  ,\nn )$.  $\CC   _{\rm
scatt}(\rr , \nn  )$ is the  contribution from scattering.

\begin{figure}\label{fig_scattering plane}
\begin{center}
\epsfig{file=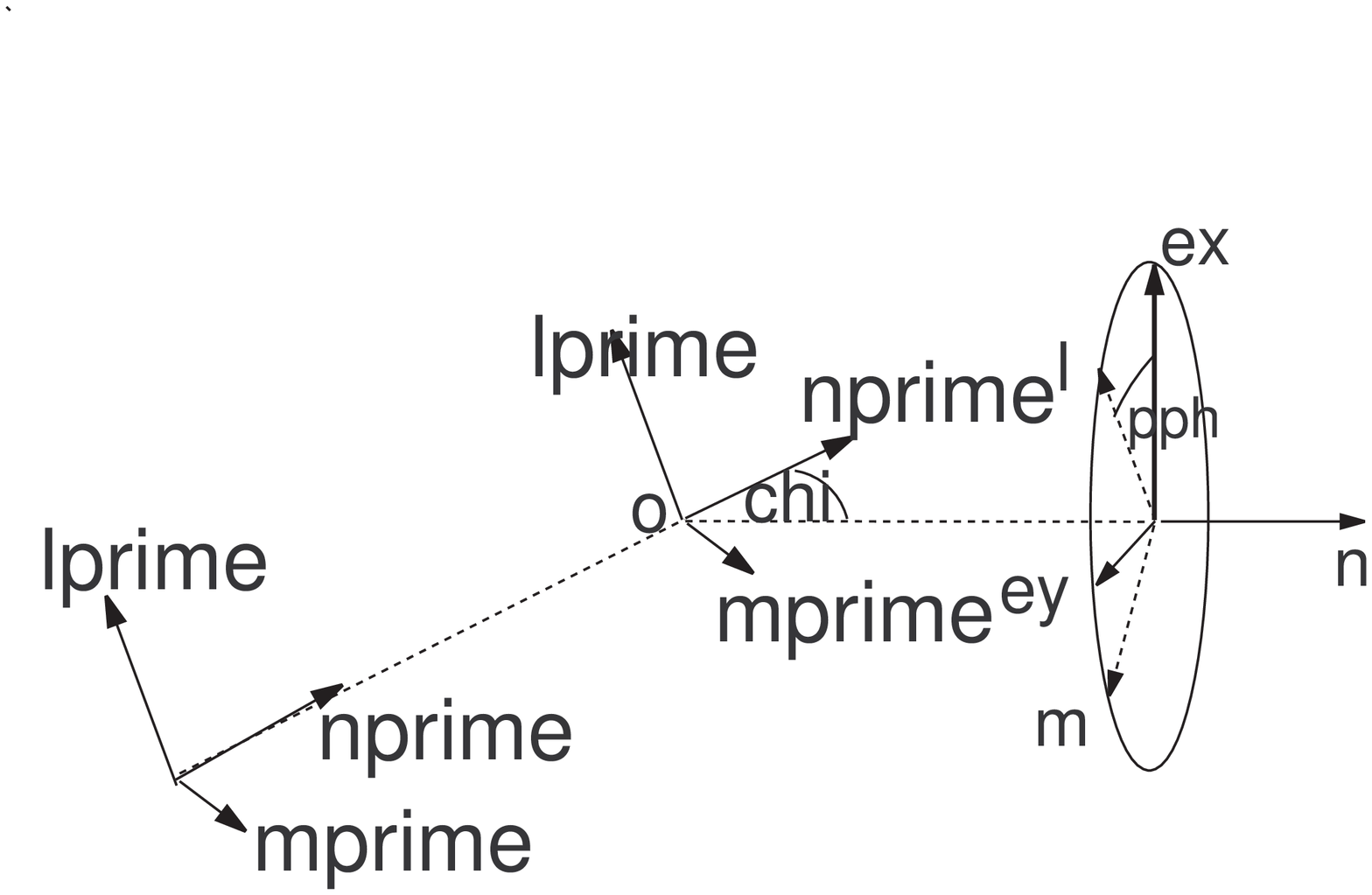, height=8cm}
\end{center}
\caption{The incident photon direction $\nn '$, and scattered photon direction,
$\nn ,$ define athe scattering plane. $\ll '$ is taken to be the normal to
the scattering plane, and $\mm '$ completes the right hand frame $\{\ll ' ,\mm ' ,\nn '\}$
for the incident photon. The scattered photon frame is defined by the right hand
frame $\{\ll  ,\mm  ,\nn \},$ where $\ll $ is taken to be equal to $\ll ' .$}
\end{figure}

If we consider a mean stellar luminosity of $L$, and a stellar number density
denoted by $\rho (\rr) $, then, for unpolarized sources $\CC _{\rm stars}(\rr ,
\nn )= (\rho (\rr ) L /4\pi, 0, 0, 0 ).$ Photons will be scattered from all
directions into  the direction $\nn $, and  so $\CC _{\rm scatt}(\rr ,
\nn )$ will depend  on the value of  the intensity in every direction,
$\nn '$, at  $\rr$. Thus we  are dealing  with an integro-differential
equation, which  can, except  for  exceptional cases, only  be  solved
numerically.  It  is natural to   consider  the solution  of  equation
(\ref{radiative  transfer})       along rays   (characteristics).  The
parametric  equation for the  ray passing through some arbitrary point
with radius vector $\rr_0$ is $\rr = \rr _0 + s\nn$,  where $s$ is the
distance of $\rr$ from $\rr_0$ along the ray. Introducing the notation
$\rr '  =  \rr _0 +   s'\nn$, equation (\ref{radiative  transfer}) now
takes the form

\begin{equation}\label{rays}
  \part {\II (\rr' , \nn )}{s} = \CC (\rr', \nn ) -\DD (\rr', \nn ) ,
\end{equation}
which has the formal solution

\begin{equation}\label{formal solution}
  \II (\rr, \nn ) = \II(\rr_0-t\nn , \nn ) +
  \int^{s}_{-t} \CC (\rr', \nn )- \DD (\rr' , \nn ) ds' ,
\end{equation}
where $-t$ is the parameter  value at the  initial point of integration along
the characteristic. Allowing $t \rightarrow \infty  $, and using the boundary
condition  $\II (\rr_0 -t\nn  ,\nn )  \rightarrow \bf {0}$  as $t \rightarrow
\infty $, equation (\ref{formal solution}) becomes

\begin{equation}\label{formal solution 2}
  \II (\rr, \nn) =  \int^{s}_{-\infty} \CC _{\rm
stars} (\rr' , \nn )ds'  + \int^{s}_{-\infty} \CC  _{\rm
scatt} (\rr' , \nn )- \DD (\rr' , \nn ) \ ds' .
\end{equation}
The term $\DD$, the extinction, is simply given by $\DD (\rr , \nn ) = n (\rr
) \sigma \II  (\rr, \nn )$, where  $\sigma $ is  the scattering cross section
and $n(\rr )$ the number density of scatterers.  $\CC _{\rm  scatt} $ is more
complicated, since it involves an integral of the Stokes intensities over all
incoming directions, $\nn  '$.   Moreover, the  scattering plane for  photons
scattered into the line  of sight $\nn $ varies  with the incident direction,
and so the total contribution has to be obtained from an appropriate rotation
to some  reference plane.  Thus   consider an incoming photon from  direction
$\nn '$ that is scattered by an electron or  dust particle at position $\rr $
into  the direction $\nn $ (see  figure \ref{fig_scattering plane})  . $\nn $
and $\nn '$ define the scattering plane, the normal to which  is given by the
vector $\nn \times\nn '$. It is natural to define  a right handed orthonormal
basis $\{\ll ' , \mm ' , \nn '\}$  associated with the incoming photon, where
$\ll ' = \nn \times \nn  '/{|\nn \times \nn  ' |}$ is  the unit normal to the
scattering  plane and $ \mm  ' = \nn'  \times \ll  ' .$ The  right hand basis
$\{\ll  ,\mm ,\nn \}$,  where $\ll=\ll '$  and  $ \mm =   \nn \times \ll $ is
associated with the scattered photon.

\psfrag{lprime}{${\bf l}'    $} \psfrag{nprime}{${\bf n }' $}
\psfrag{mprime}{${\bf m}'    $} \psfrag{l}     {${\bf  l}  $}
\psfrag{m}     {${\bf m}     $} \psfrag{n}     {${\bf  n}  $}
\psfrag{ex}    {${\bf e}_x   $} \psfrag{ey}    {${\bf e }_y$}
\psfrag{pph}   {$\tilde{\phi}$} \psfrag{chi}   {$\chi      $}

\begin{figure}\label{intensity}
\epsfig{file=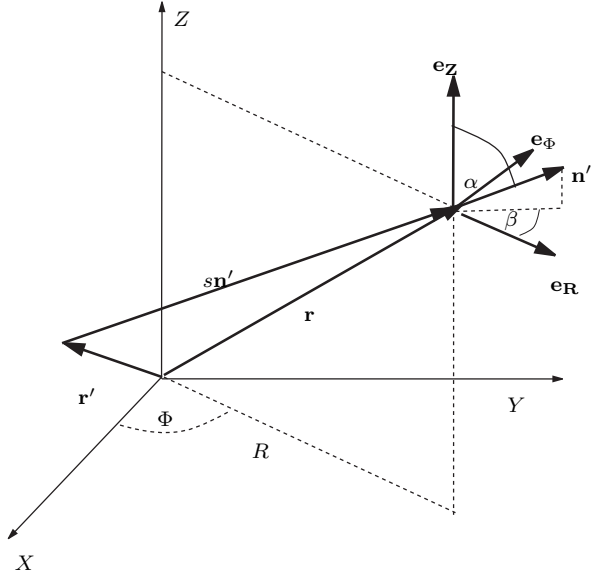, height=8cm}
\caption{The intensity in direction $\nn '$ at $rr$ due to stellar distribution
is obtained by integrating the stellar density along the line
in the direction of $\nn ' .$ The axis of symmetry is $Z .$
The polar and azimuthal angles of $\nn '$ in the
the cylindrical frame are  given by $\alpha$ and $\beta$ respectively.}
\end{figure}

We denote by $\II ' $ the Stokes parameters for incoming photons measured in the
frame $\{\ll ' ,\mm ' ,\nn '\}$ . Incoming photons in solid angle $d\Omega_{\nn
'}$ scattered into direction $\nn$ give a contribution, $d\II(\rr , \nn)$, per
unit volume to the scattered intensities in frame $\{\ll,\mm ,\nn\}$ given by
$d\II = \sigma n (\rr )\SS (\nn , \nn ') \II '(\rr , \nn ')d\Omega_{\nn '} .$
This defines the scattering phase function $\SS(\nn ,\nn ')$. To integrate the
different contributions to the scattered intensities from different incoming
rays, we need to express all the scattered Stokes intensities in the same fixed
reference frame. We take this fixed frame to be the observer's frame $\{\ee_x,
\ee_y, \ee_z\}$, which is chosen such that $\ee_z$ is the direction from the
galaxy to the observer, $\ee_x$ lies in the plane of the axis of symmetry of the
galaxy and the line of sight, and $\ee_y$ completes the right hand basis. Since
in this context we are only interested in photons scattered towards the
observer, we may without loss of generality put $\nn = \ee_z$. Thus in order to
sum the contributions from different beams we need to rotate from basis $\{\ll ,
\mm ,\nn \}$ to $\{\ee_x, \ee_y, \ee_z\}$, that is through and angle $\tilde
\phi$ given by $\cos \tilde \phi = \ee_y \cdot \ \mm .$ Under such rotations the
Stokes parameters transform as

\begin{equation}
\tilde {\II} = \RR (\tilde \phi ) \II
\end{equation}
where
\begin{equation}\label{rotation}
 \RR =
  \begin{pmatrix}
   1          &0                 &0                    &0  \\
   0          &\cos 2\tilde \phi & -\sin 2\tilde \phi  &0  \\
   0          &\sin 2\tilde \phi &\cos 2\tilde \phi    &0  \\
   0          &0                 &0                    &1
   \end{pmatrix}.
\end{equation}
Thus equation(\ref{formal solution}) becomes
\begin{equation}\label{intsol}
\begin{split}
  \II(\rr ,\nn) =  \II_0 (\rr ,\nn)
+ \sigma \int_{-\infty} ^s ds'  n(\rr ')
\int_{4\pi} \RR (\tilde\phi)\SS (\nn, \nn ')\II (\rr ' , \nn ') d\Omega_{\nn'}\\
 -\int_0 ^s n(\rr ') \sigma  \II (\rr ' , \nn) ds' ,
  \end{split}
\end{equation}
where $\II_0 (\rr ,\nn)= \int^{s}_{-\infty} \CC _{\rm stars} (\rr'
, \nn )ds'$ is the intensity at $\rr$ in the direction $\nn$
(i.e.when no scattering or absorption is present). (Any cylindrical
distribution, $\rho (R , z)$, of stars, which we assume also
radiate unpolarized radiation, must necessarily give rise to a
source field, $\II_0 (\rr ,\nn)$, that is cylindrically symmetric
and unpolarized.) We may easily express the source intensities in
terms of $\rho (R , z)$. Indeed, from figure \ref{intensity} it can
be seen that

\begin{equation}\label{source intensity}
  \II_0(\rr , \nn ) = \frac {L}{4 \pi}( \int _0 ^{\infty }\rho (\rr - s \nn
  ) d s , 0, 0, 0).
\end{equation}
It is convenient to introduce the normalised stellar density,
\begin{equation}\label{normalised density}
  \tilde{\rho}=\frac{\rho}{\int\rho d^3 \rr}.
\end{equation}
Equation \ref{source intensity} can then be written
\begin{equation}\label{source intensity2}
  \II_0(\rr , \nn ) = \frac {L_G}{4 \pi}( \int _0 ^{\infty }\tilde \rho (\rr - s \nn
  ) d s , 0, 0, 0),
\end{equation}
where $L_G$ is the galaxy's luminosity. Let us define the normalised source
intensity
\begin{equation}\label{normalised intensity}
  \Sigma (\rr , \nn )= \frac {4 \pi}{L_G}\II_0(\rr , \nn ) .
\end{equation}
$ \Sigma (\rr , \nn )$ is essentially a stellar surface density.
\psfrag{xs}{$X$}
\psfrag{ys}{$Y$}
\psfrag{zs}{$Z$}
\psfrag{np}{${\bf n}'$}
\psfrag{snp}{$s{\bf n}'$}
\psfrag{er}{${\bf e_R }$}
\psfrag{ez}{${\bf e_Z}$}
\psfrag{eph}{${\bf e_\Phi}$}
\psfrag{np}{${\bf n}'$}
\psfrag{r}{$\bf r$}
\psfrag{rp}{$\bf r '$}
\psfrag{ph}{$\Phi$}
\psfrag{rho}{$R$}
\psfrag{b}{$\beta $}
\psfrag{alp}{$\alpha $}

We  shall  consider several   scattering mechanisms. For   Thomson
scattering by electrons the scattering phase function $\SS$ takes the form

\begin{equation}
  \SS =\frac{3}{16 \pi}
   \begin{pmatrix}
     1+\cos^2 \chi & \sin^2 \chi    &0          &0 \\
     \sin^2 \chi   & 1+\cos^2 \chi  &0          &0 \\
     0             &    0           & 2\cos\chi &0 \\
     0             &    0           &0          & 2\cos\chi
\end{pmatrix}
\label{thomson scattering matrix}
\end{equation}

In the case of Rayleigh scattering (molecules and small dust particles) the form
of the scattering phase function is the same. However the total scattering cross
section behaves as $\sigma \propto \lambda ^{-4}$.

In  many numerical simulations the  Henyey--Greenstein form for scattering is
used as an approximation to a typical mix of galactic dust particles.

In this case the scattering matrix takes the form

\begin{equation}\label{Shg}
{\bf S}_{hg}  = \frac{1 }{4 \pi}
  \begin{pmatrix}
   P_1 & P_2  & 0   & 0     \\
   P_2 & P_1  & 0   & 0     \\
   0   & 0    & P_3 & -P_4  \\
   0   & 0    & P_4 &  P_3
   \end{pmatrix}.
\end{equation}

where

$$P_1 = \frac{1-g^2}{(1+g^2 - 2 \, g \, c_{\chi} )^{3/2}}$$

and $P_2$ is given by

$$\frac{P_2}{P_1} = -p_{max} \frac{1-c_{\chi}^2}{1+c_{\chi}^2} $$

$P_3$ and  $P_4$  are irrelevant  for  our  study, since we   are only
concerned with unpolarized incident  light.  The single parameter $g$,
which represents the mean value of the cosine of the scattering angle,
determines  how peaked in the  forward  direction  the scattering  is.
$p_{max}$ is the peak polarization (i.e.  the polarization at $90^o$).
For the scattering of  optical light  on dust,  $g$ and $p_{max}$  are
both of the order of $0.5$ \cite{White79}. We use these values in all 
our numerical calculations.  Throughout, we have assumed that there is
no  absorption and that   attenuation  is purely  due  to  scattering,
corresponding to the case of albedo 1.

\section{The optically thin approximation}\label{optically thin}

If the galaxy is optically thin, in the sense that $\int n \sigma dl$ through
the galaxy is less than one in all directions, then the first term on the right
hand side of equation (\ref{intsol}) will dominate. The optically thin
approximation is obtained from first order iteration, in which $\II (\rr , \nn
')$ in the integral over solid angles in the second term, and $\II_0 (\rr
,\nn ')$ in the third term are replaced by and $\II_0 (\rr ,\nn' )$ and $\II_0
(\rr ,\nn )$ respectively. Let us introduce the notation $d\tau_z = \sigma n(\rr
) dz .$ Thus in the optically thin approximation the asymptotic Stokes
intensities in the direction of the observer, $\II (x,y,\infty ,\ee_{z} )$, are
given by

\begin{equation}
\begin{split}
\II (x,y,\infty , \ee_{z}) = \II _0 {(x, y ,\infty , \ee_{z})} 
+ \int_0 ^{\tau(x,y)} d\tau_z
\int_{4\pi} \RR (\tilde\phi)\SS (\ee _z, \nn ')\II_0
{(\rr, \nn ')} d\Omega_{\nn'}\\
 -\int_0 ^{\tau(x,y)} \II_0(\rr, \ee_z) \,  d\tau_z .
\end{split}
\end{equation}
where $\tau(x,y)$ is the optical depth through the galaxy at the field point $(x,y) .$

Writing  equation (\ref{optically thin}) in   component form, and introducing
the elements $s_{ij}$ of the scattering phase function, $\SS$, we obtain

\begin{equation}\label{iintensity}
\begin{split}
  I(x,y, \infty , \ee_z) = I_{0}(x, y, \infty )
 + \int_0 ^{\tau(x,y)} d\tau_z \int_{4\pi} s_{11}(\nn, \nn ')
  I_{0}(\rr , \nn ')d\Omega_{\nn'} \\
- \int_0 ^{\tau(x,y)}
 I_{0}(\rr , \ee_z ) d\tau_z
\end{split}
\end{equation}

\begin{equation}\label{qintensity}
  Q(x,y, \infty , \ee_z) =
  \int d\tau_z
  \int_{4\pi}d\Omega_{\nn'} I_{0}(\rr , \nn ')(s_{21}(\nn, \nn ')
  \cos 2\tilde\phi - s_{31}(\nn, \nn ')\sin 2\tilde\phi )
\end{equation}

\begin{equation}\label{uintensity}
  U(x,y, \infty , \ee_z) =
  \int d\tau_z
  \int_{4\pi} d\Omega_{\nn'}I_{0}(\rr , \nn ')(s_{21}(\nn, \nn ')\sin 2\tilde\phi
  + s_{31}(\nn, \nn ')\cos 2\tilde\phi)
\end{equation}
and
\begin{equation}\label{vintensity}
  V(x,y, \infty , \ee_z) = \int d\tau_z
  \int_{4\pi}d\Omega_{\nn'} I_{0}(\rr , \nn ') s_{41}(\nn, \nn ')
\end{equation}
where $I_{0}$ is given in terms of the stellar number density by equation
(\ref{source intensity}). Introducing the notation

\begin{equation}
  W = Q + i U
\end{equation}
we obtain
\begin{equation}\label{complex}
W (x,y, \infty , \ee_z) =
  \int d\tau_z
  \int_{4\pi} d\Omega_{\nn'}I_{0}(\rr , \nn ')\exp
  {i2\tilde\phi}((s_{21}(\nn, \nn ')
  + i s_{31}(\nn, \nn ')).
\end{equation}

In the case of spherical scattering, where the angular dependence of $s_{ij}$
is only in the scattering angle, $\theta$, one can expand $s_{11}$ ($s_{41} =
0$ in this  case) in terms  of  Legendre polynomials $P_l$,  and $s_{21}$ and
$s_{31}$ in   terms of associated  Legendre polynomials,  $P_{lm}$  with $m =
2$. This leads to simple expressions for the asymptotic Stokes intensities in
terms of  the line integrals of the  multipole moments of the $\II_{0}(\rr ,
\nn )$, with $m = 0$ for the $I$ and $V$ and $m = 2$ for $Z$. This is similar
to the results obtained by \cite{Simmons82} and \cite{Simmons83} for the point
source case. For Thomson and Rayleigh scattering further simplification results
from the fact that only terms with $l = 0, 2$ occur in the expansion of the
phase functions $s_{ij}$. This has a crucial importance when we come to
calculate the flux. We shall not pursue this line of reasoning here, but rather
adopt the simpler approach adapted to the case of Thomson and Rayleigh
scattering. However, we outline the method using spherical harmonics section
\ref{harmonics}.

\subsection{Polarization map}\label{spm}

Using equations \ref{iintensity}--\ref{vintensity}, it is possible to compute
polarization maps of  the  galaxy.  To  carry out  this we  have  divided the
galaxy field into  $100 \times 100$ pixels, and  for each each pixel we  have
integrated along  the  line of   sight.   The results  our   shown in  figure
(\ref{pol_carte}), and can be seen to be comparable with those obtained
by Wood  \cite{Wood97a}, who used  Monte  Carlo  techniques.   Of course  the
integration  is in our  case extremely quick.   The degree of polarization is
much higher for a galaxy viewed edge on than for the  same galaxy viewed face
on, and this is essentially due to the respective  optical depths. One should
note that for Henyey--Greenstein scattering  the upper and  lower half of the
galaxy polarization  is no longer  symmetric.  This is  due to the functional
form of the scattering, which is now peaked in the forward direction.

One can  obtain the polarized flux   from the polarization intensity  maps by
integrating the  contributions to  the polarization  over the field  of view.
However,  as  we show  in the  sections \ref{thin  flux} and \ref{harmonics},
analytic  expressions  for the  polarized flux  in  the  case of Thomson  and
Rayleigh scattering can be   found, and approximate expressions  for  general
spherical scattering mechanisms.

\setlength{\unitlength}{1cm}
\begin{figure}\label{pol_carte}
\begin{picture}(18,13)
\put(-1.0,0.0){
\psfig{file=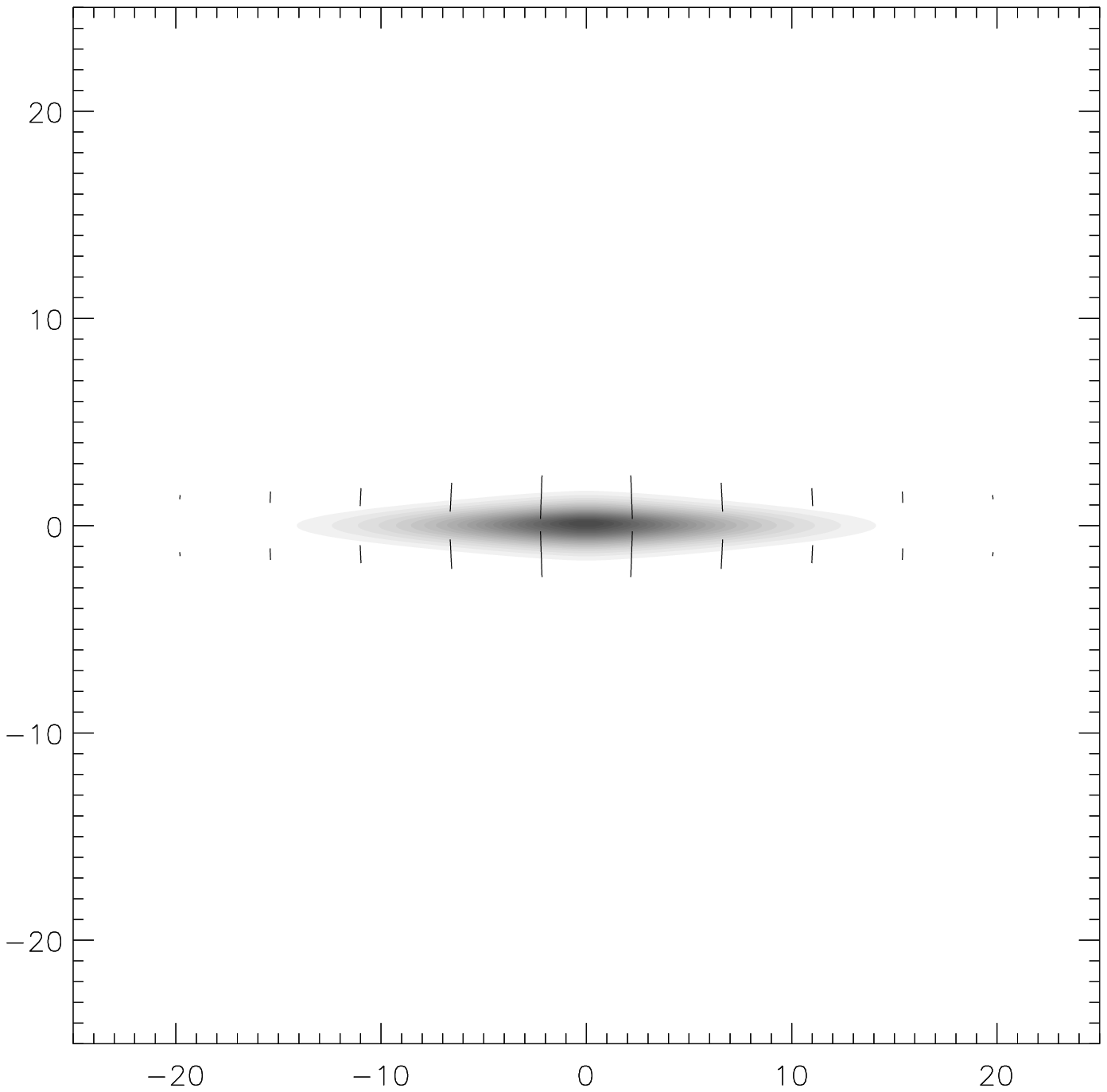 ,height=7.0cm,width=7.0cm}}
\put(4.75,0.0){
\psfig{file=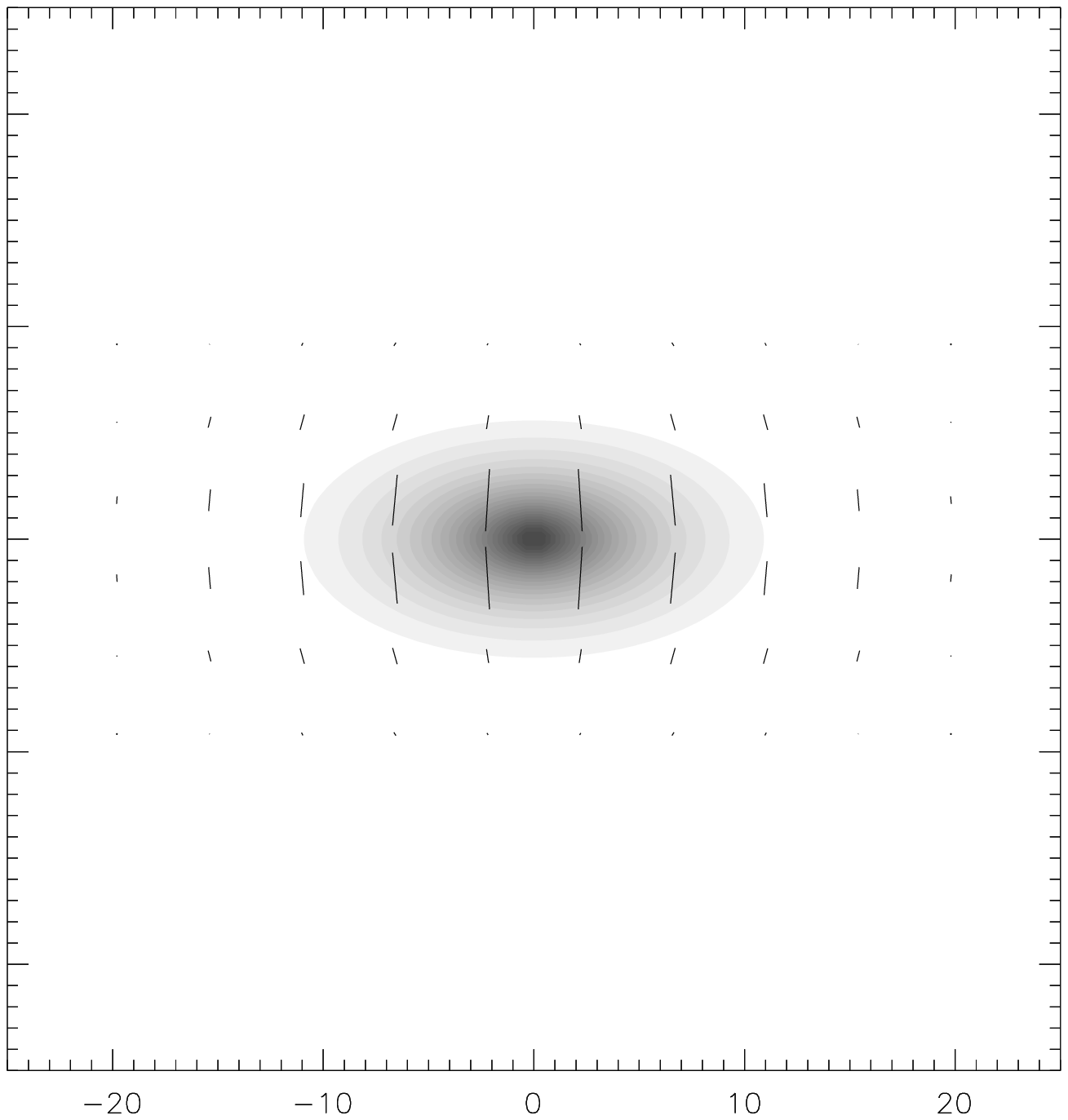 ,height=7.0cm,width=7.0cm}}
\put(10.5,0.0){
\psfig{file=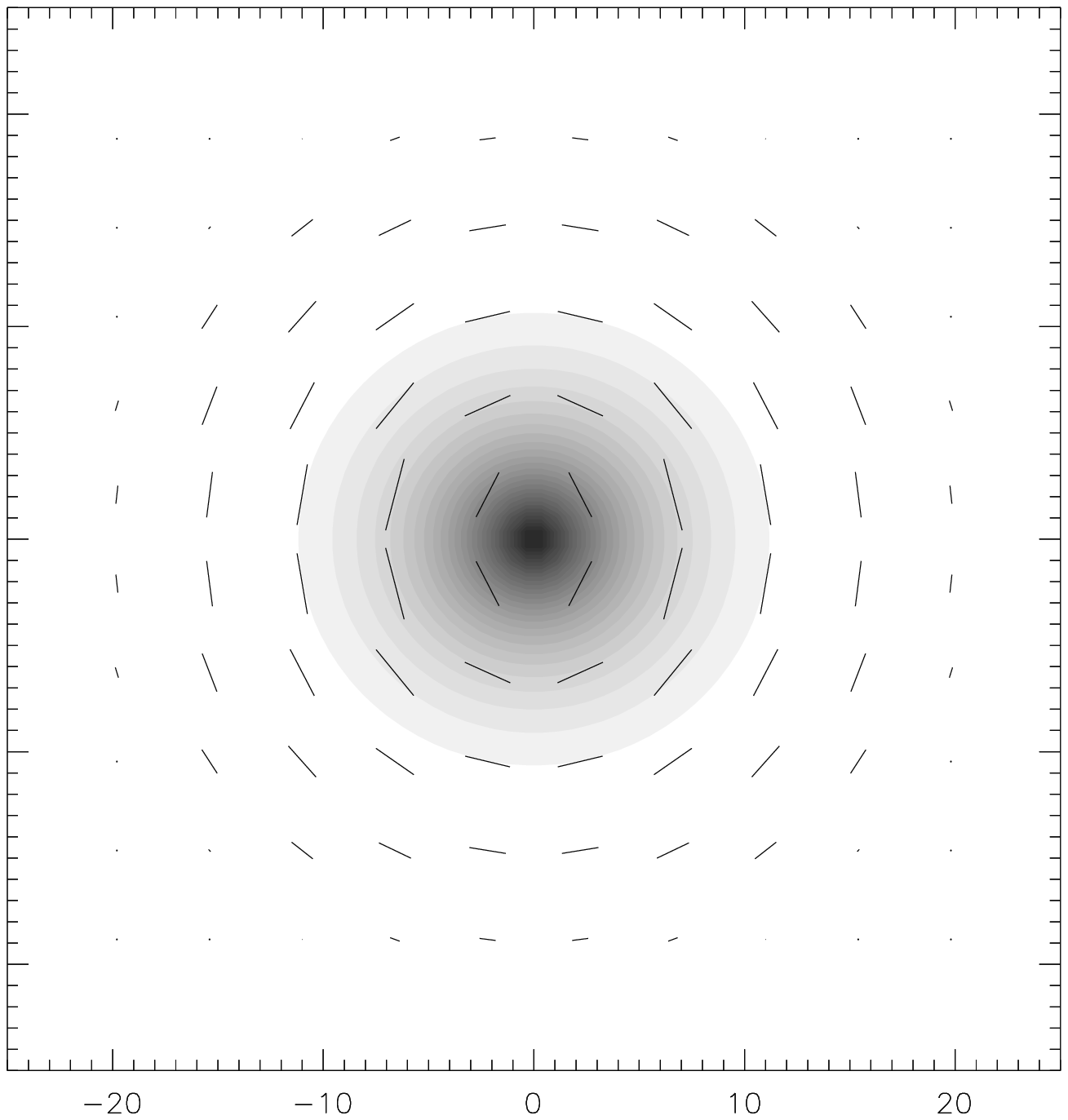 ,height=7.0cm,width=7.0cm}}
\put(-1.0,6.03){
\psfig{file=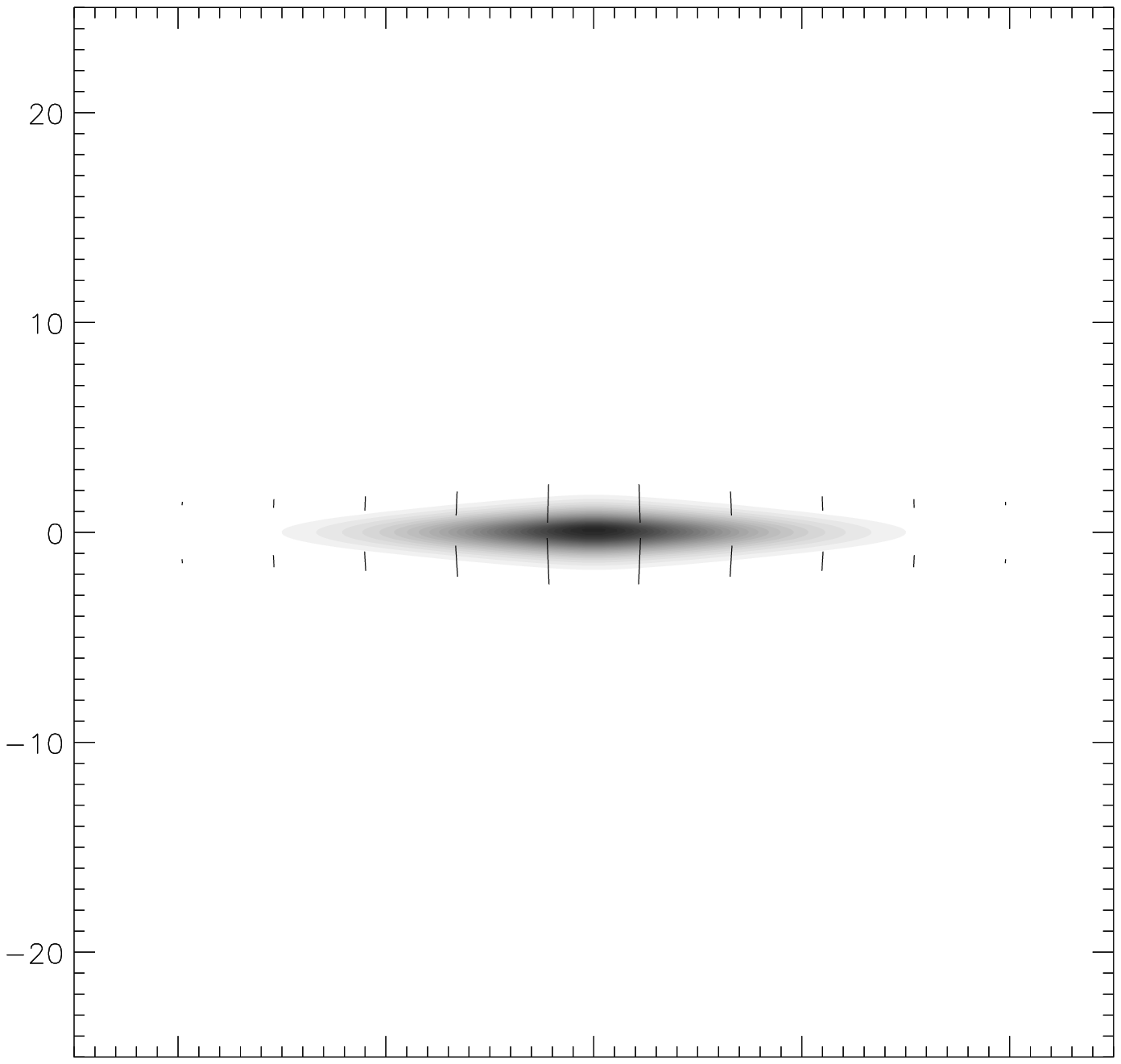 ,height=7.0cm,width=7.0cm}}
\put(4.75,6.03){
\psfig{file=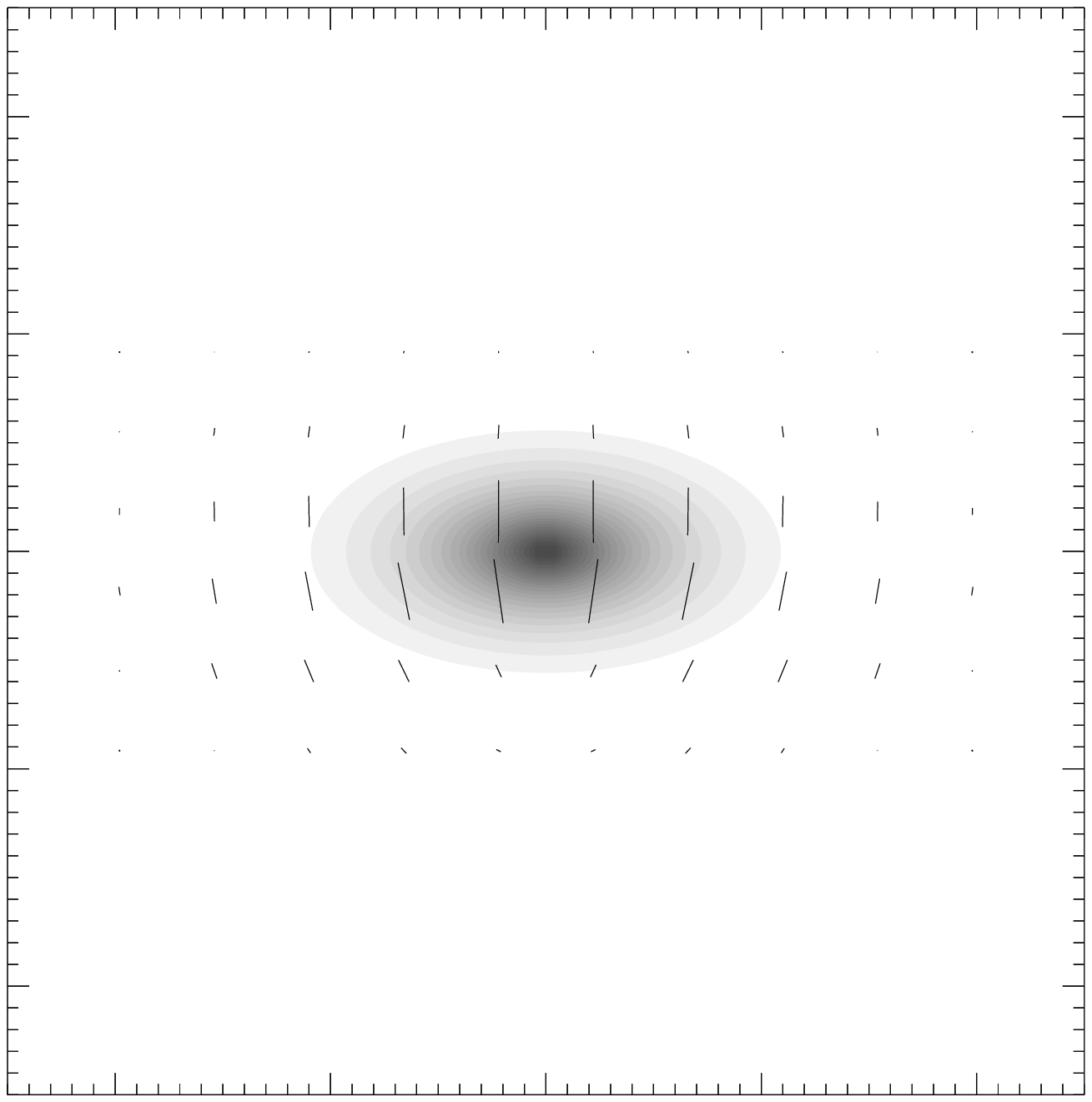 ,height=7.0cm,width=7.0cm}}
\put(10.5,6.03){
\psfig{file=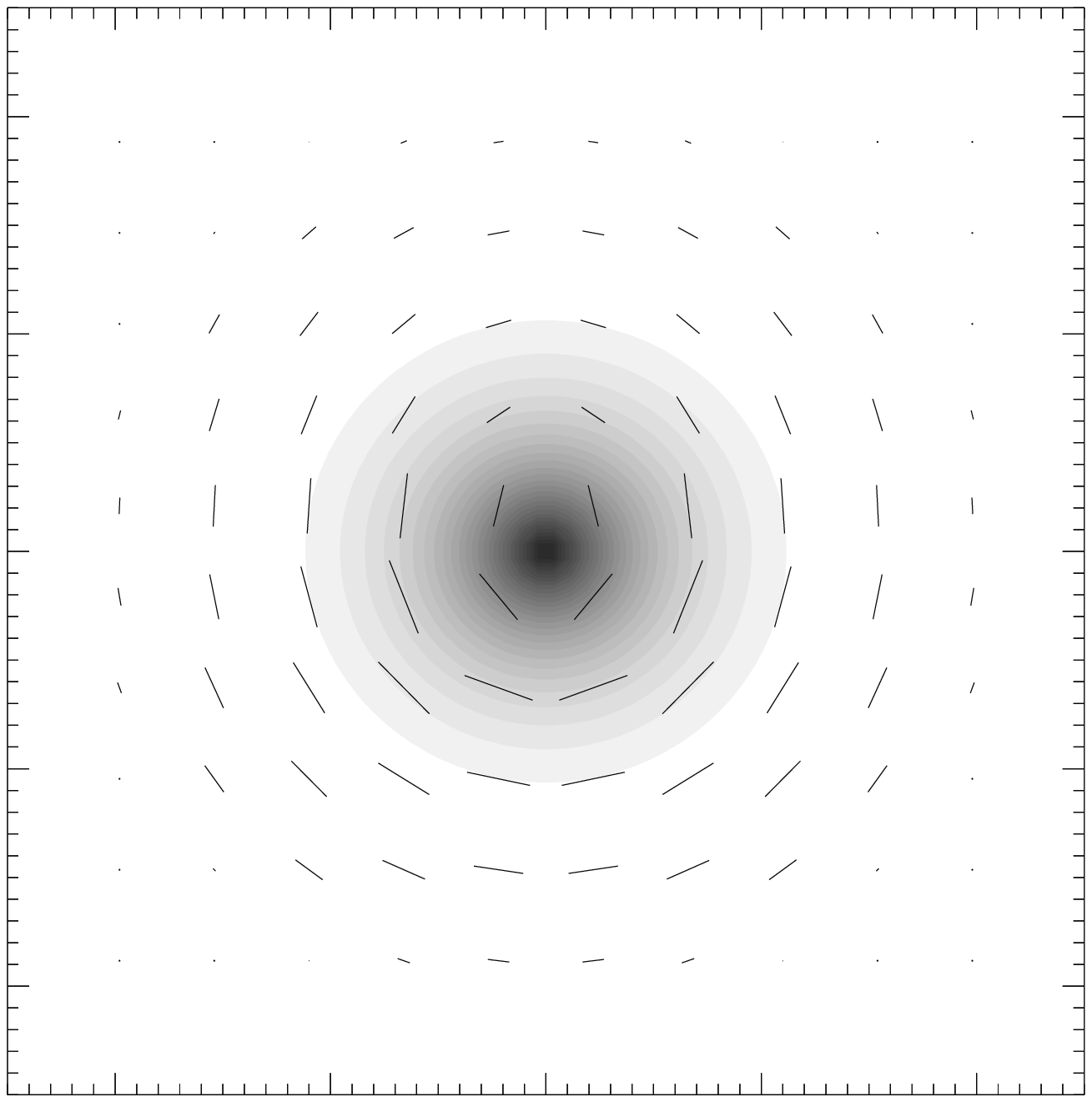 ,height=7.0cm,width=7.0cm}}
\end{picture}
\caption{Polarisation  maps   for  Thomson  scattering  (bottom   plots)  and
Henyey--Greenstein scattering (top plots). The inclinations of the galaxy are
$87^o$, $60^o$ and $18^o$ for the left, central and right plots respectively.
The degree of polarization  is proportional to  the length of the  lines, the
maximum polarization being 0.8\% , 0.5\% and 0.2\% for  the top row and 1.8\%,
1.1\% and 0.4\% for the bottom row. The axis a labelled in $kpc$.}
\end{figure}

\begin{figure}\label{prof_pol}
\begin{center}
\epsfig{file=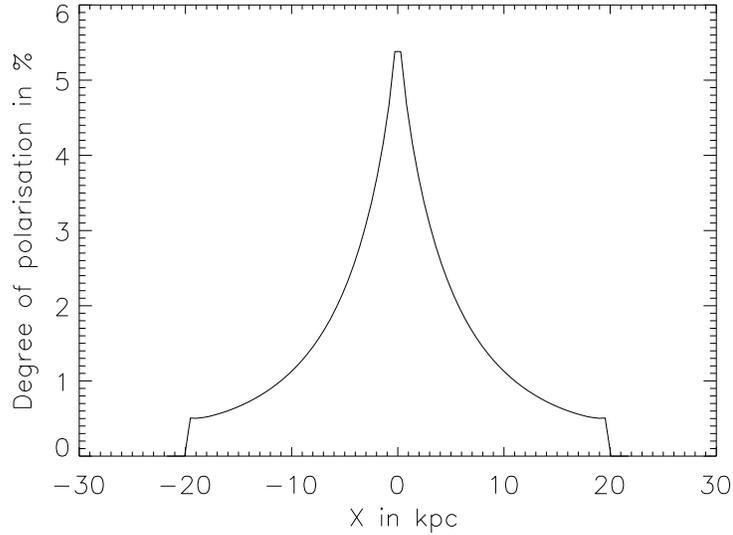, height=8cm}
\end{center}
\caption{Polarization profile  along the major axis  of a galaxy  for Thomson
scattering for an inclination $87^o$. For Henyey--Greenstein the profile is
very similar but the overall degree of polarization lower. The cut-off at $\pm 20 kpc$
corresponds to the cut-off we have applied in the disc distribution, but the
intensity at this distance from the galactic centre is already extremely low}
\end{figure}

\subsection{Polarized flux for Thomson and Rayleigh scattering}
\label{thin flux}

The flux at the earth is obtained  by integrating the asymptotic intensities,
i.e. $\II(x,y,\infty ,\ee_z)$, for large $s$ over all solid angles as seen by
the observer. Noting that this solid angle is just given by  $dx dy / R_E ^2$
it is clear that  integration of  equation  (\ref{intsol}) over solid  angles
gives an expression for the flux in terms of volume  integrals and the direct
flux, $\FF_0 (\infty )$.

\begin{equation}\label{fluxsol}
\begin{split}
 \FF = \frac {1}{R_E ^2}  \int \II (x,y,\infty ,\ee_z) dx dy  =
 \FF_0 +
   \int d V  \sigma n(\rr ) \int \RR (\tilde\phi)\SS (\ee_z , \nn ')
  \II_0 (\rr , \nn ') d\Omega_{\nn'}\\
  -\int \sigma n(\rr) \II_0 (\rr , \ee_z ) dV.\\
  \end{split}
\end{equation}
From equation (\ref{source intensity2}) it follows that
\begin{equation}\label{source flux}
 \FF_0 (\infty ) =\frac{L_G}{4\pi  R_E ^2}( 1 , 0, 0, 0 )
\end{equation}

When calculating the degree of polarization to first order in optical depth, we
need only normalize $F_U$ and $ F_Q $ by the direct unpolarized flux, ignoring
the contribution to the unpolarized flux from scattering.

\begin{figure}\label{fig_scattering geometry}
\epsfig{file=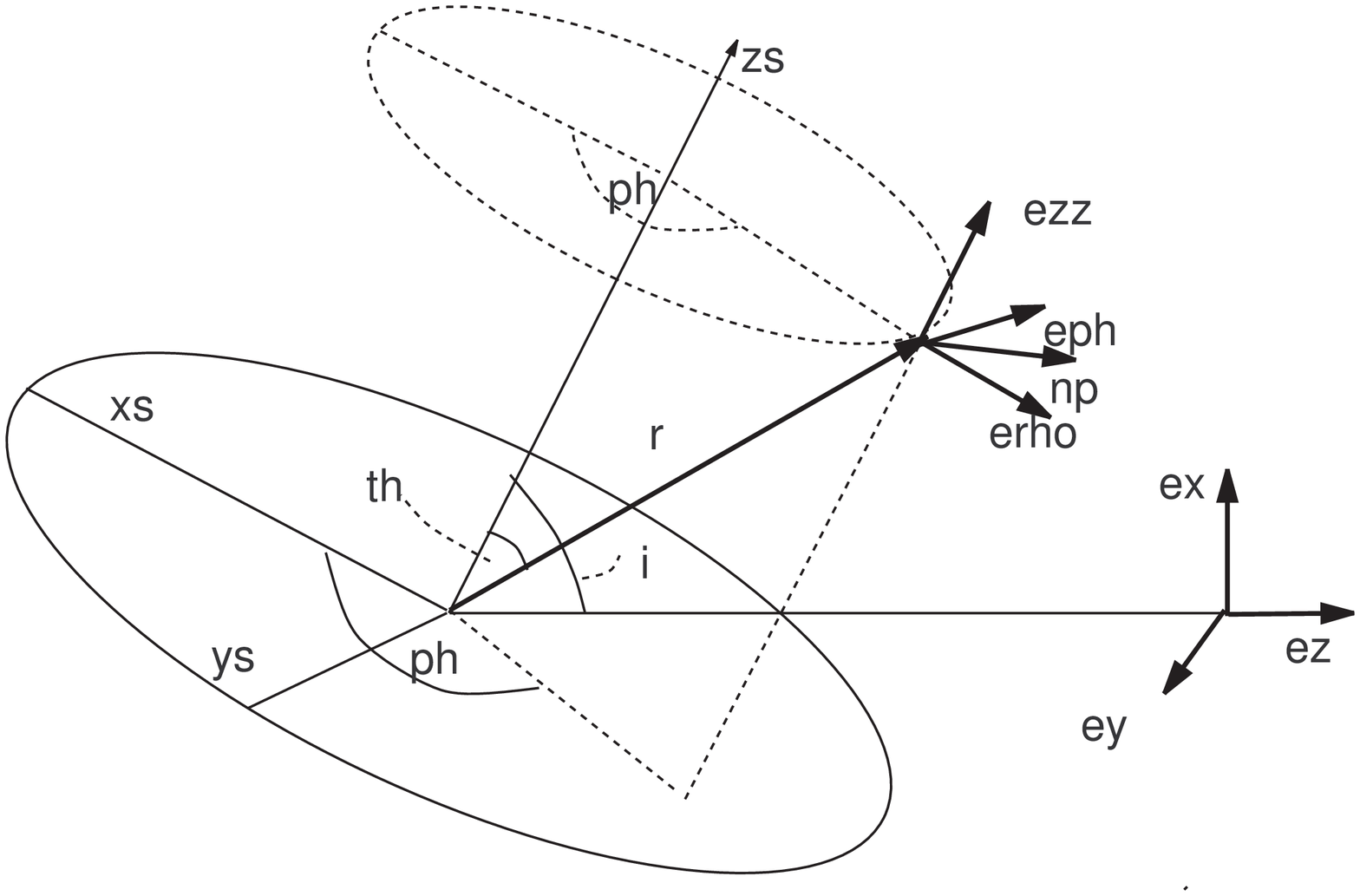, height=8cm}
\caption{The axis of symmetry, given by $Z$, is inclined at angle $i$ to the line of sight.
The observer's frame is $\{\ee_x ,\ee_y ,\ee_z \}$.
$OX$, $OZ$ and $\ee_x$ are taken to lie in the same plane.}
\end{figure}

Let us now assume that the galaxy has axial symmetry. Thus the distribution of
stars and the density distribution of scatterers are taken to be cylindrically
symmetric about this axis. The axis of symmetry is inclined at an angle of $i$
to the line of sight. It is useful to introduce cylindrical coordinates with
origin at the centre of the galaxy, $\{ R , \Phi ,Z \} $ , with the $Z$
direction oriented along the axis of symmetry of the galaxy. (See figure
\ref{fig_scattering geometry} .)
\psfrag{xs}{$X$}
\psfrag{ys}{$Y$}
\psfrag{zs}{$Z$}
\psfrag{nprime}{${\bf n}'$}
\psfrag{mprime}{${\bf m}'$}
\psfrag{erho}{${\bf e}_ R $}
\psfrag{ezz}{${\bf e}_{Z}$}
\psfrag{eph}{${\bf e}_\Phi $}
\psfrag{np}{${\bf n}'$}
\psfrag{i}{$i$}
\psfrag{r}{$\bf r$}
\psfrag{ex}{${\bf e}_x$}
\psfrag{ey}{${\bf e}_y$}
\psfrag{ez}{${\bf e}_z$}
\psfrag{ph}{$\Phi$}
\psfrag{th}{$\Theta$}

It is also convenient to write $n = n_0 \; \tilde n ,$ where $n_0$ is the
central density of scatterers, and $ \tilde n $ thus a dimensionless density. We
introduce further a length scale $Z_g$ and the dimensionless variable $Z' =
Z/Z_g$, which will enable us to express the Stokes fluxes in terms of the
optical depth, $\tau_0 = n_0 \sigma Z_g $. This is indeed the vertical optical
depth for the models of section \ref{mod_gal}. We shall also work with the
normalised intensity, $\Sigma (\rr , \nn ) ,$ given by equation \ref{normalised
intensity}.

Because of the symmetry, the density of scatterers, $\tilde n$, is only be a
function of $R$ and $Z$. Indeed this is the form assumed in the models of
section \ref{mod_gal}. Similarly the unpolarized source Stokes intensity,
$\Sigma (\rr , \nn ) ,$ is a function of $R$ and $Z$, and the direction cosines,
$\alpha , \beta$ of $\nn '$, expressed in terms of the associated coordinated
basis vectors $\{\ee_R ,\ee_\Phi ,\ee_Z \}$.

We shall now prove that under these assumptions, and with the additional
assumption that the dust is optically thin, $F_Q$ depends only on $\sin^2 i$,
and $F_U = 0$, generalizing the result obtained \cite{Brown77} for point
sources. The behaviour of the degree of polarization will differ slightly from
this, since the normalization factor, $F_I$, also depends weakly on the
inclination. However, to first order in the optical depth the result will still
hold. We also obtain an explicit form for the polarized flux in terms of density
moments.  Equation (\ref{fluxsol}) written in
component form now becomes

\begin{equation}\label{thinfluxIsol}
\begin{split}
 F_I = F_0 (\infty ) + \frac{L_G}{4 \pi}\frac{3  \tau_0 }{16 \pi R_E ^2}
                        \int  \tilde n (R , Z )  R dR dZ' d\phi \int (1+\cos ^2 \chi )
                        \ \Sigma (R , z, \alpha , \beta )
                         \ \sin \alpha \, d\alpha \, d\beta \\
 -\int  n (R , Z ) \sigma I_0 (R , Z, \alpha , \beta ) dV\\
  \end{split}
\end{equation}

\begin{equation}\label{thinfluxQsol}
\begin{split}
 F_Q = \frac{L_G}{4 \pi}\frac{3 \tau_0 }{16 \pi R_E ^2}
       \int  \tilde n (R , Z ) R  dR dZ' d\phi \int  \sin ^2 \chi \;
      \Sigma (R , Z, \alpha , \beta ) \cos 2\tilde\phi \ \sin \alpha \, d\alpha \, d\beta\\
 \end{split}
\end{equation}

and

\begin{equation}\label{thinfluxUsol}
\begin{split}
 F_U = \frac{L_G}{4 \pi}\frac{3\tau_0}{16 \pi R_E ^2}
                        \int \tilde n (R , Z )R dR  dZ'  \  d\phi \int  \sin ^2 \chi \;
                        I_0 (R , Z, \alpha , \beta ) \sin 2\tilde\phi
                        \ \sin \alpha \, d\alpha \,d\beta .
\end{split}
\end{equation}
To  evaluate $F_I , F_Q$ and  $F_U$ we shall need  to work out the scattering
angle $\chi$  (between  $\nn = \ee_z$ and  $\nn  '$) and the  rotation  angle
$\tilde\phi$  between the  scattering plane  and the  observer's $y$ axis  in
terms of $  \Phi , \alpha  ,  \beta$. (As we  shall see,  although we  do not
require an explicit expression  for the scattering  angle in terms of $\Phi ,
\alpha , \beta$ for the calculation of $F_Q$ and $F_U$, for the evaluation of
$F_I$ we shall require it.)

In order to  do this we need  to express the  basis $\{ \ee_{R }, \ee_{\Phi},
\ee_{Z} \}$ in terms of $\{\ee_x  , \ee_y , \ee_z \}$.   Let us introduce the
convenient shorthand notation for cosine, $\cos  A = c_A$,  and sine, $\sin A
=s_A$.

\begin{equation}\label{basis transformation}
\begin{pmatrix}
 \ee_{R} \\
 \ee_{\Phi}  \\
 \ee_{Z}   \\
\end{pmatrix}
 =
  \begin{pmatrix}
  c_\Phi              & s_\Phi                &0  \\
   -s_\Phi             & c_\Phi                & 0 \\
   0                      &0                         & 1 \\
\end{pmatrix}
 \begin{pmatrix}
   c_i                       &0                          & s_i   \\
    0                      & 1                            & 0 \\
   -s_i                      &0                         & c_i \\
\end{pmatrix}
\begin{pmatrix}
 \ee_{x} \\
 \ee_{y}\\
 \ee_{z}\\
\end{pmatrix}.
\end{equation}
The first matrix on the right  hand side represents  a transformation from
$\{ \ee_{R }, \ee_{\Phi}, \ee_{Z} \}$  to $\{ \ee_{X}, \ee_{Y}, \ee_{Z} \}$,
and the second  from  $\{  \ee_{X}, \ee_{Y},  \ee_{Z}  \}$  to
$\{  \ee_{x}, \ee_{y}, \ee_{z} \}$.

Equation (\ref{basis transformation}) simplifies to
\begin{equation}\label{basis transformation 2}
\begin{pmatrix}
 \ee_{R} \\
 \ee_{\Phi}  \\
 \ee_{Z}   \\
\end{pmatrix}
 =
\begin{pmatrix}
  c_\Phi c_i             & s_\Phi                &-c_\Phi s_i  \\
  -s_\Phi c_i            & c_\Phi                & s_\Phi s_i  \\
  s_i                    &0                      &c_i          \\
\end{pmatrix}
\begin{pmatrix}
 \ee_{x} \\
 \ee_{y}\\
 \ee_{z}\\
\end{pmatrix}.
\end{equation}
Writing
\begin{equation}
\nn ' =
\begin{pmatrix}
  s_\alpha c_\beta       & s_\alpha s_\beta          &c_\alpha \\
\end{pmatrix}
\begin{pmatrix}
 \ee_{R} \\
 \ee_{\Phi}  \\
 \ee_{Z}   \\
\end{pmatrix},
\end{equation}
and using equation (\ref{basis transformation 2})
we obtain
\begin{equation}\label{nprimematrix}
\begin{split}
  \nn ' = (s_\alpha  c_\beta  c_\Phi  c_i - s_\alpha  s_\beta  s_\Phi c_i + c_\alpha
  s_i)\ee_{x} \\
  + (s_\alpha  c_\beta  s_\Phi     +    s_\alpha  s_\beta  c_\Phi ) \ee_{y} \\
  + (s_\alpha  c_\beta  c_\Phi s_i +    s_\alpha s_\beta  s_\Phi s_i + c_\alpha  c_i)\ee_{z} \\
\end{split}
\end{equation}
Noting that $\cos\tilde \phi = (\nn ' \times \ee_z)\cdot \ee_y /|\nn ' \times
\ee_z |$ and $\sin  \tilde \phi  = (\nn '   \times \ee_z)\cdot \ee_x  /|\nn '
\times \ee_z |$ we immediately obtain from equation (\ref{nprimematrix})
\begin{equation}
   \label{cosine rotation}
  \cos\tilde \phi = -\frac {c_\alpha s_i + s_\alpha c_\beta c_\Phi c_i
  - s_\Phi c_i s_\alpha s_\beta }
  {s_\chi}
\end{equation}
and
\begin{equation}
   \label{sine rotation}
  \sin \tilde \phi = \frac {s_\alpha c_\beta  s_\Phi
  + s_\alpha s_\beta c_\Phi}
{s_\chi} .
\end{equation}
The scattering angle, $\chi$, is given by $\cos  \chi = (\nn ' \cdot \ee_z)$,
and again from equation (\ref{nprimematrix}) we obtain
\begin{equation}\label{scattering angle}
  c_\chi = c_\alpha  c_i  -
  s_\alpha c_\beta  c_\Phi s_i + s_\alpha s_\Phi s_\beta  s_i
  .
\end{equation}
Noting that $\cos 2\tilde\phi = \cos^2 \tilde\phi - \sin^2 \tilde\phi$, and
$\sin 2\tilde\phi = 2\cos \tilde\phi \sin \tilde\phi$ and using eqs.
(\ref{cosine rotation}) and (\ref{sine rotation}), we see that in the integrand
of equations \ref{thinfluxQsol} and \ref{thinfluxUsol} the terms $ \sin ^2 \chi
\cos 2\tilde\phi$ and $ \sin ^2 \chi \sin 2\tilde\phi$ simplify by virtue of the
cancellation of the $\sin ^2 \chi$ terms, and yield respectively
\begin{equation}\label{simplify}
\begin{split}
s^2 _{\chi}   c_{2\tilde\phi} = (c_\alpha  s_i
         +s_\alpha c_\beta  c_\Phi c_i
        - s_\Phi c_i s_\alpha s_\beta )^2  -(
        s_\alpha c_\beta  s_\Phi
        + s_\alpha s_\beta c_\Phi )^2 ,
\end{split}
\end{equation}
and
\begin{equation}
  \begin{split}
  s^2_\chi  s_{2\tilde\phi} = -2(c_\alpha s_i
  +s_\alpha c_\beta  c_\Phi c_i
  - s_\Phi c_i s_\alpha s_\beta )
  (s_\alpha c_\beta s_\Phi
  + s_\alpha s_\beta c_\Phi ).
  \end{split}
\end{equation}

Upon carrying out the integration over $\Phi$ all terms in $\sin \Phi$, $\cos
\Phi$,  and  $\sin  2\Phi$   disappear,  since both  $n$   and  $I$ are
independent of  $\Phi $. A calculation shows  that only terms with the factor
$\sin ^2 i$   remain in the final  expression  for $F_Q$,  and  $F_U =0$. The
latter implies  that the  polarization  is in the  y  direction or  in  the x
direction,  i.e. along or   perpendicular to plane   defined by  the axis  of
symmetry and the line of sight.  Explicitly we obtain
\begin{equation}\label{polarized flux}
\begin{split}
F_Q =\frac{L_G}{4 \pi} \frac{\tau \sin ^2 i}{16  R_E ^2} &
                        \int \tilde n (R , Z ) R dR dZ'   \int  I_0 (R , Z,  \alpha , \beta)
                       ( 3c_\alpha^2
                        -1) \ s_\alpha \, d\beta d\alpha .
\end{split}
\end{equation}
Equation (\ref{polarized flux}) has  a  simple interpretation: the  polarized
flux is proportional  to the dipole  moment of  the radiation  field weighted
over the  density distribution of the scatterers.

A similar calculation can be carried out for the unpolarized flux.
Substituting in equation (\ref{thinfluxIsol}) the expression for $\cos \chi$
given in equation (\ref{scattering angle}), we obtain

\begin{equation}\label{flux}
\begin{split}
F_I = F_0 (\infty) +
\frac{L_G}{4 \pi}\frac{3\tau_0 \sin ^2 i}{16  R_E ^2}
                        \int \tilde n (R , Z ) R \ dR \ dZ'   \int  I_0 (R , Z,  \alpha , \beta)
                       ( 2 + 2 c_\alpha^2 c_i ^2 - s_\alpha ^2 s_i ^2
                       c_{2\beta} )
                        \ s_\alpha d\beta d\alpha \\
                        -\int n(R , Z ) \sigma  I_0 (R , Z, \alpha , \beta )\  dV
                        .
\end{split}
\end{equation}

The degree of linear polarization, $p$, is given, to first order in the optical
depth, as $F_Q / F_0 (\infty)$. From equations (\ref{polarized flux})and
(\ref{source flux}) we obtain

\begin{equation}\label{degree of polarization}
  p = \frac{3\pi \tau_0 \sin ^2 i}{16  }
                        \int R dR dZ'  n(R , Z ) \int  \Sigma (R , Z,  \alpha , \beta)
                       ( 3c_\alpha^2
                        -1) \ s_\alpha d\beta d\alpha
\end{equation}
where $\Sigma (R , Z,  \alpha , \beta)$ is given by equation \ref{normalised intensity}.

\subsection{Approximate expressions for polarized flux using spherical harmonics}
\label{harmonics}

For more general mechanisms than Thomson and Rayleigh scattering the analytic
methods  used in section  \ref{thin flux} are not  very helpful. The approach
used there worked because of the cancellation of the  factor $\sin^2 \chi$ in
the integrand  of the polarized fluxes owing  to  the form of  the scattering
phase function. For most phase functions this will not be the case.

Spherical harmonics,  $Y_{lm}(\theta   , \phi )$,   provide us   with  a very
powerful method of dealing  with  the general  case.  On  the one  hand  they
provide a complete  set of  orthogonal functions on   the sphere, and  on the
other for  each value of  $l $ the $2l+1$ function  $Y_{lm}(\theta  , \phi )$
provide an irreducible representation of the rotation group. We shall largely
follow the notation of Messiah (1961), and take the spherical harmonics to be
defined by
\begin{equation}\label{eee}
 Y_{lm}(\theta , \phi ) = c(l,m) \ P_{l}^ {m} (\cos \theta )e^{im\phi}
 \     \    {\rm where} \   \ c(l,m) = [ \frac {(2l+1)(l-m)!}{4\pi
 (l+m)!}]^{1/2}
\end{equation}
and $P_{l}^  {m}$  are the  associated  Legendre polynomials.  $c(l,m)$  is a
normalisation constant to ensure $Y_{lm}(\theta , \phi )$ are orthonormal.

Thus suppose that we have two coordinate bases, $\{\ee_x, \ee_y, \ee_z\}$ and
$\{\ee_x ',  \ee_y ' , \ee_z '\}$.  Suppose further  that the rotation taking
$\{\ee_x, \ee_y, \ee_z\}$ into $\{ \ee_x ', \ee_y  ' , \ee_z '\}$ is described
by the Euler angles $A , B , C$.  We then have the relation
\begin{equation}\label{spherical harmonics}
Y_{lm}(\theta ' , \phi ' ) = \sum_{m' =-l}^{m' =l} Y_{l m' }(\theta  , \phi  )
R^{(l)}_{m'm} (A , B ,C  ).
\end{equation}
The matrices $\RR ^{(l)}$ form  an irreducible representation of the rotation
group. The elements of $\RR ^{(l)}$ take the form
\begin{equation}\label{representation}
R^{(l)}_{m m'} (A , B ,C  ) = e^{-im A}r^{(l)}_{m m'} (B  )e^{-im'C},
\end{equation}
where $r^{(l)}_{m'm} (B)  $ can be calculated from the Wigner formula (see Messiah 1961),
\begin{equation}\label{Wigner}
r^{(l)}_{m m'} = \sum_t (-1)^t \frac {\sqrt{(l+m)!(l-m)!(l+m')!(l-m')!}}
{(l+m-t)!(l-m'-t)!t!(t-m+m')!}\xi ^{2l+m-m'-2t}\eta ^{2t-m+m'},
\end{equation}
where $\xi = \cos B/2$ and $\eta = \sin B/2$.

It is convenient for our purposes to work entirely in the observer's reference
frame (see figure (\ref{fig_scattering geometry}). Our two frames are thus $\{
\ee_x, \ee_y, \ee_z \}$, the observer's frame, and $\{\ee_x ', \ee_y ' , \ee_z
'\}=\{ \ee_R, \ee_\Phi,
\ee_Z \}$, the frame attached to the galaxy's cylindrical coordinates.
With this notation we replace $\theta '$ by $\alpha$, and $\phi '$ by $\beta .$
To rotate the coordinate basis $\{ \ee_x, \ee_y,
\ee_z \}$ into $\{ \ee_R, \ee_\Phi, \ee_Z \}$ we have to

(i) rotate about $y$-axis though $i$ and then

(ii)rotate about $Z$-axis though $\Phi $. Thus $A=0$, $B=i$, and $C=\Phi .$

Let us assume that the source Stokes intensity has cylindrical symmetry, i.e.
$I_0 = I_0 (R , Z , \alpha , \beta)$.  Let us work in terms of the normalised
surface density $\Sigma (R , Z , \alpha , \beta) ={4\pi}I_0 (R , Z , \alpha ,
\beta)/{L_G}$ as defined in equation \ref{normalised intensity}.
Expanding $ \Sigma (R , Z , \alpha , \beta)$ in terms of spherical harmonics
\begin{equation}\label{harmonic expansion}
\Sigma  (R, Z , \alpha , \beta) =
\sum_{l=0}^{l=\infty} \sum_{m=-l}^{m=l}a_{lm}(R, Z )Y_{lm }(\alpha , \beta)
\end{equation}
where of course $a_{lm}$ do not depend on $\Phi $. Indeed because of the
orthogonality of the spherical harmonics we may write
\begin{equation}\label{harmonic coefficient}
  a_{lm}(R, Z ) = \int_{4\pi} \Sigma  (R, Z , \alpha , \beta) Y_{lm }^* (\alpha ,
  \beta)d\Omega .
\end{equation}
Now expressing $Y_{lm }(\alpha , \beta ) $ in terms of $Y_{lm }(\theta , \phi )$
via equation \ref{spherical harmonics}, and substituting into equation
\ref{harmonic expansion}, we obtain
\begin{equation}\label{harmonic expansion 2}
\Sigma  (R , Z , \alpha , \beta) =\sum_{l=0}^{l=\infty} \sum_{m=-l}^{m=l}\sum_{m'=-l}^{m'=l}
a_{lm}(R , Z ) Y_{lm '}(\theta  , \phi  ) R^{(l)}_{m'm} (0 , i  , \Phi  ).
\end{equation}
Now from equation {representation}, $R^{(l)}_{m'm} (0 , i , \Phi ) =
r^{(l)}_{m'm} ( i ) \exp {-i m \Phi }$.

The  angle ( denoted  by $\tilde  \phi$  in figure \ref{fig_scattering plane})
between the scattering plane of the incoming photon  in direction $\nn $, and
the  observer's  $x-z$ plane, is $\tilde   \phi = \phi   -  \pi /2$,  and the
scattering angle, $\chi$ in the main text, is $\chi = \theta$.

The complex intensity, $W$, is given  by equation \ref{complex} . Note
that for  spherical   scattering mechanisms   we need  only   consider
$s_{21}$ and  moreover   $s_{21}  =  s_{21}(\theta)$. To  obtain   the
polarized flux,  expression (\ref{complex}) for the complex intensity,
$W =  Q + i  U$, has to  be integrated  over all  solid  angles at the
observer. As in  section (\ref{thin flux})  we multiply by the element
of solid   angle, $dx dy  /R_E  ^2$, thus transforming  the expression
(\ref{complex}) into a volume integral, where we have written $d\tau_z
= \sigma n_0 \tilde n dz$ .

As before, introducing the dimensionless variable $Z' = Z/Z_g $, and the central
optical depth, $d\tau_0 $, we thus obtain the expression
\begin{equation}\label{complex flux}
  F_Q + iF_U = -\frac{\tau_0 L_G }{ 4\pi R_E ^2}\int \tilde n (R , Z) R  dR d Z'
  d\Phi \int s_{21}(\theta) \Sigma  (R , Z , \alpha , \beta )
  e^{i 2 \phi } \sin \theta d\theta d\phi .
\end{equation}
Substituting  equation  (\ref{harmonic expansion  2}) into  equation (\ref
{complex flux}) we obtain
\begin{equation}\label{complex flux 2}
  \begin{split}
  F_Q + iF_U =
  -\frac{\tau_0 L_G }{4\pi R_E ^2}\int \tilde n(R , Z) R  dR \  dZ'
  d\Phi \int s_{21}(\theta)
  \sum_{l=0}^{l=\infty} \sum_{m=-l}^{m=l}\sum_{m'=-l}^{m'=l}\\
  a_{lm}(R , Z )
 Y_{lm '}(\theta  , \phi  )
r^{(l)}_{m'm} ( i   ) e^{-i m \Phi} e^{i 2 \phi } \sin \theta d\theta d\phi
\end{split}
\end{equation}
Integration  over  $\Phi$ eliminates all  terms  for   which $m\neq  0$,  and
introduces a factor $2\pi $ for $m=0$. With the definition
\begin{equation}\label{Alm}
  A_{lm} = \int  \tilde n (R , Z) \, a_{lm}(R , Z )\, R  \, dR \, dZ'
\end{equation}
equation (\ref{complex flux 2}) reduces to
\begin{equation}\label{complex flux 3}
\begin{split}
F_Q + iF_U =
  -\frac{L_G \tau_0 }{2 R_E ^2}
  \sum_{l=0}^{l=\infty} A_{l0}\sum_{m'=-l}^{m'=l}
\int s_{21}(\theta)
 Y_{lm '}(\theta  , \phi  )
r^{(l)}_{m'0} ( i   )  e^{i 2 \phi } \sin \theta d\theta d\phi .
\end{split}
\end{equation}
Clearly, only the  term $m' = -2$  can contribute, and integral (\ref{complex
flux 3}) reduces to
\begin{equation}\label{multipoles}
  F_Q + iF_U = -\frac{L_G  \tau_0 }{ 2 R_E ^2}
  \sum_{l=2}^{l=\infty} A_{l0} r^{(l)}_{-2 0} (i)S_{l2\  (21)}
\end{equation}
where
\begin{equation}\label{scattering harmonic}
 S_{l2\ (21)}= 2\pi  \ c(l, -2)\int s_{21}(\theta )P_{l\,2}(\cos \theta
)\sin \theta d\theta .
\end{equation}

Since all the  terms on the right hand  side equation  (\ref{multipoles}) are
real, it follows that $F_U$ is zero,  meaning that the polarization is either
directed  along the  axis  of symmetry,  or perpendicular to  it. For Thomson
scattering the only  non-zero term in  the  multipole expansion of  the phase
function is $S_{22 \ (21)}$,  and we  obtain  the result proved in  \ref{thin
flux}, since $r^{(2)}_{-2 0}(i)$ is proportional to  $\sin^2 i$. On the other
hand,  if  the  scattering phase   function has higher   multipoles, then the
dependence of the polarization  on  the inclination  angle will have   higher
terms in $\sin i $ and $\cos i$.

To first order in optical depth, the degree of polarization is given by, $p =
|F_Q|/F_0(\infty)$.  Noting that $F_0(\infty) = L_G  / 4\pi R_E ^2$ we obtain
for the degree of polarization
\begin{equation}\label{final polarization}
p = 2\pi \tau_0 \sum_{l=2}^{l=\infty} A_{l0} \; r^{(l)}_{-2 0} (i)S_{l2}
\end{equation}
A similar calculation can be carried out for $F_I$.

\subsection{Dependence of polarized flux on inclination}

\begin{figure}
\label{pol_ang}
\begin{center}
\epsfig{file=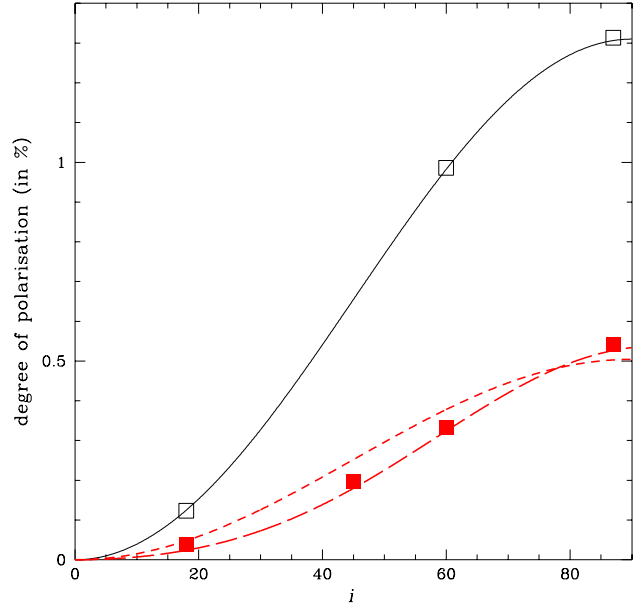, height=8cm}
\end{center}
\caption{Degree of polarization as  a function of  $i$.  The empty squares
show the numerical  values for  Thomson scattering  and the full  squares for
Henyey--Greenstein  scattering.   The   $sin^2$  law  is,  as   was expected,
recovered     for the   Thomson   scattering    (solid   line)  .   For   the
Henyey--Greenstein   scattering there is    a noticeable deviation  from this
law. The  short-dashed line  corresponds  to the  l=2 approximation, and  the
long-dashed line to the l=4 approximation for Henyey-Greenstein scattering.}
\end{figure}

From the polarization  maps presented in  section (\ref{spm}), it is possible
to  calculate   the integrated   degree  of   polarization over  the   entire
galaxy. This  must be a function  of the inclination. The corresponding plots
of polarization (calculated in   this way) against  inclination are  given in
figure (\ref{pol_ang}).   For Thomson scattering one obtains  the $\sin ^2 i$
law, as expected from the analytic results of section \ref{thin flux}. On the
other hand, for  Henyey--Greenstein    scattering,  there is a     noticeable
deviation from this law.

The dependence  of   the total   polarization on   inclination can  also   be
calculated using   the  expansion  in spherical   harmonics  outlined  in the
previous section.  The degree of polarization is given by equation \ref{final
polarization}.

$A_{l0}$ is an  integral over the distribution   of stars, dust and  gas, and
does not depend on the scattering mechanisms or inclination. $r^{(l)}_{-2 0}$
is a function  of inclination that is easy  to calculate.   $S_{l2\ (21)}$ is
determined from the scattering   function from the  simple integral  given by
equation \ref{scattering harmonic}.Results for $l=2, 3$ and $ 4$ are shown in
the table \ref{tab1}.  For Thomson  scattering only $l=2$  is non-zero.   For
Henyey--Greenstein higher order multipoles are not  zero, nevertheless, as is
illustrated in figure (\ref{pol_ang}), even with inclusion  only of the $l=2$
term a  very good fit is found.  Including terms up to  $l=4$, one obtains an
excellent  fit to the numerical  results, with a  disagreement  of only a few
percent.

Evidently this method can test an  arbitrarily
large number of different scattering mechanisms and geometries extremely
quickly, since the total polarization is given by a simple sum of products of
easily calculated terms. Thus to see the effect of changing the scattering
mechanism, it is sufficient to simply recalculate the corresponding
coefficients $S_{l2\ (21)} .$

\begin{table*}
\begin{center}
\begin{tabular}{|c|c|c|c|c|}
  &   $ r^{(l)}_{-20} $     &  $ A'_{l0}$  & $S_{l2\ (21)}$ (thom.) &
$S_{l2\ (21)}$ (HG)\\ \hline\hline
l=2  & $\frac{\sqrt{24}}{8} \sin^2(i)$ &  $ 2.21 \;10^{-2}$   &  $\sqrt{3 \pi/10}$  &
$0.373$       \\ \hline
l=3  & $\frac{\sqrt{30}}{4} \sin^2(i) \cos(i)$&  $-4.16 \;10^{-3}$  &  $0.$           &
$0.174$       \\ \hline
l=4  & $\frac{\sqrt{10}}{16} \sin^2(i) (5 + 7 \cos(2i))$&  $-1.72\;10^{-2}$  &  $0.$     &
$0.043$      \\ \hline
\end{tabular}
\caption{   }
\label{tab1}
\end{center}
\end{table*}

\section{Discussion and conclusions}
\label{conclusions}

The  assumption that the spiral  galaxy is optical  thin allows  us to easily
calculate   numerically  the spatially  resolved   degree of polarization and
unpolarized intensity of starlight scattered by dust,  electrons and gas. Our
numerical   results agree  qualitatively   with the  observations  for spiral
galaxies, and  for  vertical optical depths  of  around  $0.05$  give maximum
polarization of around 1\% depending on the dominant scattering mechanism and
inclination of the galaxy.  For the same  optical depth, Thomson and Rayleigh
scattering  are more  efficient       in producing polarization  than     the
Henyey/Greenstein  phase function.  The  latter  produces an  asymmetry in the
polarization about the semi-major axis of the elliptical image of the galaxy.

Although we should  expect the optical thin assumption  might break down  for
high  inclination galaxies,  which  are indeed the  ones  discussed  by \cite
{Draper95}  and \cite {Scarrott96},  and where  the  optical depth  along the
galactic plane is greater than $1$, our results appear to agree well with the
Monte Carlo calculations of a  number of authors.  It would be interesting to
make a detailed comparison of the two methods.

The advantages of the optical thin assumption  is undoubtedly its efficiency:
the computer time needed for calculation of maps considerably less than Monte
Carlo    treatment.  Although  we    have   dealt only   with  Thomson    and
Henyey--Greenstein type  scattering functions, the   inclusion of mixtures  of
scatterers etc. would be very straightforward, and involve only slightly more
computer time.
It would is  a  straightforward extension to treat   a case where  the
galaxy is  optically  thick  in  absorption,   but optically  thin  in
scattering.  However  in  this case  we  should  not  expect the  same
inclination law to hold for the polarization. We  shall deal with this
case in a future paper.
There have been  only few observational  studies  of the polarization  of the
integrated  flux from spiral   galaxies. We find  that in  the optically thin
regime,  for Thomson or Rayleigh scattering,  the dependence of the degree of
polarization on inclination of  the galactic axis  to the line of sight, $i$,
has a  simple   $\sin^2 i$  form  if we  assume  axial symmetry   in both the
distribution of  scatterers and stars. Again, for  vertical optical depths of
$\tau  \approx 0.05$, the total  integrated  polarization reaches about $1$\%
for higher inclinations.

Thus the suggestion that optical polarization could be used in the study of weak
lensing \cite{Audit99} is further supported by this study. Although the
observation of such levels of polarization would be difficult in such high
redshift galaxies, the more important determination of the direction of
polarization might be feasible.

Such a technique, in determining the orientation
of the source galaxy, would provide considerably more precision in determining
the mass distribution in weak lensing studies.

For more complicated  scattering mechanisms one  can write down an expression
for   the total  integrated polarization  in  terms of  a  spherical harmonic
expansion expansion. For the class of galaxy models we have taken, it appears
that the convergence of  this expansion  is very  rapid,  and that to a  good
approximation the polarization is  given by the first  term of the expansion,
$l=2$.   This indicates that  the  $\sin^2 i$  law  should  be more generally
applicable. Our  formulation easily allows  the inclusion of other mechanisms
(absorption and dichroism etc.).

Although we  have  not investigated the  the  joint distribution of flux  and
polarized flux for a catalogue of spiral galaxies, it would be interesting to
do so. In the optically thin regime we would expect a correlation between the
two, whereas in the optically thick regime the flux  should be independent of
inclination. Thus this could provide a test for optical thickness for a class
of spiral  galaxies. The form of the  joint distribution  could also indicate
whether scattering  or  other  mechanisms just  as  alignment  of  grains  by
magnetic  fields is    responsible  for polarization   in a   class of spiral
galaxies.

Finally we would like  to point out  the possible  use  of polarization in  a
Faber-Jackson type distance estimators arising  from the possible correlation
between polarization and absolute magnitude  of galaxies, which could be used
to refine the distance scale for galaxies.

\end{document}